\documentclass[preprint,aps,prc,showpacs,showkeys,preprintnumbers,floatfix,superscriptaddress,nofootinbib]{revtex4}
\usepackage{amssymb}

\usepackage{bm}
\usepackage{dcolumn}
\usepackage{graphicx}
\usepackage{amsmath}
\usepackage{bbold}
\usepackage[dvips]{epsfig}

\input{tcilatex}

\begin{document}

\title{ \textbf{Nucleon Properties in the Quantized Linear Sigma Model at
Finite Temperature and Chemical Potential}}
\author{H. M. Mansour}
\affiliation{Department of Physics, Faculty of Science, Cairo University, Egypt}
\author{M. Abu-Shady}
\affiliation{Department of Applied Mathematics, Faculty of Science, Menoufia University,
Egypt}
\date{\today}

\begin{abstract}
The linear sigma model at finite temperature and chemical potential is
systematically studied using the coherent-pair approximation, in which fully
taking quantum of fields are included. The expectation value of the chiral
Hamiltonian density is minimized and the resulting equations for the nucleon
are solved. The qualitative features of the quantized sigma and pion fields
are strong sensitive to the change of temperature and chemical potential and
are in agreement with the mean-field approximation calculations. It is
noticed that the nucleon mass increases with increasing coherence parameter $%
x$. In addition, the nucleon mass increases with increasing temperature $T$
\ and the baryonic chemical potential $\mu$ and then it decreases at higher
values of the temperature and baryonic chemical potential. The obtained
results show that the mean-square radius of the proton and the neutron
increase with \ increasing temperature\ or the baryonic chemical potential
and that the pion-nucleon coupling constant $g_{\pi NN}$ decreases with the
temperature or baryonic chemical potential. We conclude that the
coherent-pair approximation successfully give better description of the
nucleon properties at finite temperature and baryonic chemical potential.
\end{abstract}

\maketitle

% % % % % % % % % % % % % % % % % % % % % % % % % % %

\section{Introduction}

% % % % % % % % % % % % % % % % % % % % % % % % % %
The topic of hot and dense medium is of great interest in the high energy
heavy ion collisions $\left[ 1,2\right] $. The properties of vacuum at zero
temperature and density have different behavior at finite temperature and
density due to the chiral symmetry restoration and deconfinement of the
system.\ This gives more insight into the complex vacuum structure of
quantum chromodynamic (QCD) theory which describes relevant features of
particle physics in the early universe and the neutron stars.

Such investigation unfortunately cannot be calculated directly in the
framework of QCD theory. The difficulties involved in obtaining low-energy
properties directly from QCD, the fundamental theory of strong interactions
is due to its coupling constant at low energy scale and thereby the
necessity of dealing non-perturbatively with its complicated structure $%
\left[ 3\right] $. Hence, the investigations are usually performed by using
the effective models which share the same properties with QCD theory such as
the Nambu-Jona-Lasinio (NJL) model $\left[ 4\right] $ and the linear sigma
model and its modifications $\left[ 5-11\right] $.

At finite temperature and density, the linear sigma model and its
modifications show a significant success in the description of chiral phase
transition and the static properties of the nucleon such as discussed in
Refs. $\left[ 12-19\right] $. There are some limited research works on the
nucleon properties at finite temperature and density in the framework of the
linear sigma model. Christov et al. $\left[ 12\right] $ studied the
modification of baryon properties at finite temperature and density. They
ignored quantum fluctuations in their calculations. Dominguez et al. $[13]$
calculated the pion-nucleon coupling constant and the mean square radius of
the proton as functions of the temperature. Thermal fluctuations are only
considered and they ignored quantum fluctuations in their model. In Ref.$%
\left[ 14\right] $, the nucleon properties are also studied in the framework
of the linear sigma model at finite temperature without including chemical
potential in their calculations.

On the other hand, in Refs. $\left[ 15-17\right] $, the authors focus on the
study of the phase transition and critical point at finite temperature and
chemical potential in the framework of the linear sigma model. Bilic and
Nikolic $\left[ 15\right] $ studied the chiral transition in the linear
sigma model at finite temperature and chemical potential in mean-field
approximation. Phat and Thu $\left[ 18\right] $ studied the chiral phase
transition and its order at finite temperature and isospin chemical
potential using the linear sigma model. Schaefer et al. $\left[ 19\right] $
have extended the quark sigma model to include certain aspects of the gluon
dynamics via the Polyakov loop at finite temperature and quark chemical
potential. Mao et al. $\left[ 20\right] $ studied the deconfinment phase
transition using the Friedberg-Lee model at finite temperature and chemical
potential, and then they obtained the critical value of the temperature and
chemical potential. In the NJL model $\left[ 21\right] $, the meson and
nucleon properties are studied. they considered the case of a quark medium
as well as nucleon medium. They studied the behavior of the nucleon mass and
energy of the soliton at different temperatures and densities in the
mean-field approximation. Cheng-Fu et al. $\left[ 22\right] $ studied some
nucleon properties such as a binding energy using the simplified version of
Faddeev equations at finite temperature and chemical potential.

In the present work, we apply the coherent-pair approximation (CPA) in the
chiral quark-sigma model at finite temperature and chemical potential. CPA
is a powerful nonperturbative method to go beyond the mean-field
approximation by fully taking thermal and quantum fluctuations into account $%
\left[ 14\right] $. While CPA has already been applied in the chiral quark
sigma model at finite temperature only in Ref. $\left[ 14\right] $, so far
no attempt has been made to include the finite chemical potential. We
continue further study of the investigation that started in Ref. $\left[ 14%
\right] $. Here, our aim is to study how the nucleon properties respond when
the chemical potential is included in the framework of the linear sigma
model using CPA.

In Sec. 2, the effective mesonic potential at finite temperature and the
chemical potential is explained. The variational principle is presented in
Sec. 3. The derived nucleon properties are calculated at finite temperature
and chemical potential in Sec. 4. The numerical calculations and the results
are presented in Sec. 5. A summary and conclusions are introduced in Sec. 6.
% % % % % % % % % % % % % % % % % % % % % % % % % % % % % % % % % % % % % % %

\section{The Quark Sigma Model at finite Temperature and Chemical Potential}

% % % % % % % % % % % % % % % % % % % % % % % % % % % % % % % % % % % % % % % %
The effective mesonic potential describes the interactions of quarks via $%
\sigma$ and $\mathbf{\pi}$-meson at finite temperature $T$ and baryonic
potential $\mu$ $\left[ 17\right] $ as follows:%
\begin{align}
U_{1}^{eff}\left( \sigma,\mathbf{\pi}\right) & =U^{(0)}(\sigma ,\mathbf{\pi)}%
-12\int\frac{d^{3}P}{(2\pi)^{3}}[\sqrt{P^{2}+M^{2}}+  \notag \\
& T\ln\left( \exp\left( \frac{\mu}{T}-\frac{1}{T}\sqrt{P^{2}+M^{2}}\right)
+1\right) +  \notag \\
& +T\ln\left( \exp\left( -\frac{\mu}{T}-\frac{1}{T}\sqrt{P^{2}+M^{2}}\right)
+1\right) ].   \tag{1}
\end{align}
where, the original mesonic potential at zero temperature and chemical
potential is given by
\begin{equation}
U^{(0)}\left( \sigma,\mathbf{\pi}\right) =\frac{\lambda^{2}}{4}\left(
\sigma^{2}+\mathbf{\pi}^{2}-\nu^{2}\right) ^{2}+m_{\pi}^{2}f_{\pi}\sigma.
\tag{2}
\end{equation}
In Eq. (2), the $\sigma$ and $\pi$ are the sigma and pion meson fields. The
parameters $\lambda^{2}$ and $\upsilon^{2}$ are related to the pion and
sigma masses, and the pion decay constant $f_{\pi}=93$ MeV%
\begin{equation}
\lambda^{2}=\frac{m_{\sigma}^{2}-m_{\pi}^{2}}{2f_{\pi}^{2}},   \tag{3}
\end{equation}
\begin{equation}
\nu^{2}=f_{\pi}^{2}-\frac{m_{\pi}^{2}}{\lambda^{2}}.   \tag{4}
\end{equation}
In Eq. (1), the divergent second term comes from the negative energy states
\{$E=\sqrt{P^{2}+M^{2}}\}$of the Dirac sea. It can be partly absorbed in the
coupling constant $\lambda^{2}$ and the constant $\nu^{2}$ by using a
renormalization procedure $\left[ 16\right] $. The constituent quark mass is
defined as $M^{2}=g^{2}\left( \sigma^{2}+\mathbf{\pi}^{2}\right) $. In Eq.
(1), the second term can not be evaluated in closed form as in normally the
case for most integrals in statistical mechanics. So we expand the effective
potential in the powers of $M^{2}$ which ensures that the potential
satisfies the chiral symmetry when $m_{\pi}\rightarrow0$. \ Thus, the
effective potential takes the following form in the one-loop approximation.%
\begin{equation}
U_{2}^{eff}(\sigma,\mathbf{\pi})=U^{(0)}(\sigma,\mathbf{\pi)-}\frac{6}{\pi
^{2}}(\frac{7\pi^{4}}{180}T^{4}+\frac{\pi^{2}}{6}T^{2}\mu^{2}+\frac{1}{12}%
\mu^{4})+6g^{2}(\frac{T^{2}}{12}+\frac{\mu^{2}}{4\pi^{2}})\left( \sigma ^{2}+%
\mathbf{\pi}^{2}\right) .   \tag{5}
\end{equation}
In Refs. $[23,24]$, the authors calculated the critical temperature at zero
chemical potential by using one-loop technique in the linear sigma model.
They found that two loops or higher-order loops did not affect the critical
temperature and the phase transition. They found that two loops lead to
complex terms that are ignored in the calculation of the critical
temperature. At finite temperature and chemical potential, the one-loop
approximation is successfully predicted the chiral phase transition $\left[
25\right] .$

Now, we can write the Hamiltonian of quark sigma model at finite temperature
and chemical potential including quantized fields as in Refs. $\left[
6\rightarrow 8\right] $

\begin{equation*}
\hat{H}\left( r\right) =\frac{1}{2}\{\hat{P}_{\sigma}\left( r\right)
^{2}+(\nabla\hat{\sigma}\left( r\right) )^{2}+\hat{P}_{\pi}\left( r\right)
^{2}+(\mathbf{\nabla}\pi\left( r\right) )^{2}\}+U_{2}^{eff}(\hat{\sigma },%
\mathbf{\hat{\pi}})+
\end{equation*}%
\begin{equation}
\overset{}{\hat{\Psi}^{\dagger}}\left( r\right) (-i\alpha\mathbf{\nabla })%
\overset{}{\hat{\Psi}}\left( r\right) -g\left( r\right) \hat{\Psi }%
^{\dagger}(r)(\beta\hat{\sigma}\left( r\right) +i\beta\gamma_{5}\mathbf{\hat{%
\tau}}.\mathbf{\hat{\pi}})\hat{\Psi}\left( r\right) ,   \tag{6}
\end{equation}
where $\alpha$ and $\beta$ are the usual Dirac matrices. In the above
expression, $\overset{\symbol{94}}{\overline{\Psi}},\hat{\sigma},\ $and$\
\mathbf{\hat{\pi}}$ are quantized field operators with the appropriate
static angular momentum expansion $\left[ 7\right] ,$%
\begin{align}
\hat{\sigma}\left( r\right) & =\int_{0}^{\infty}\frac{dkk^{2}}{\left[
2\left( 2\pi\right) ^{3}W_{\sigma}\left( k\right) \right] ^{\frac{1}{2}}}%
\left[ \hat{c}^{\dagger}\left( k\right) e^{-ik.r}+\hat{c}\left( k\right)
e^{+ik.r}\right] ,  \tag{7} \\
\mathbf{\hat{\pi}}\left( r\right) & =\left[ \frac{2}{\pi}\right] ^{\frac{1}{2%
}}\int\limits_{0}^{\infty}dkk^{2}\left[ \frac{1}{2W_{\pi}\left( k\right) }%
\right] ^{\frac{1}{2}}\sum\limits_{lmw}j_{l}\left( kr\right)
Y_{lm}^{\ast}\left( \Omega_{r}\right) [\hat{a}_{lm}^{1w\dagger}\left(
k\right)  \notag \\
& +\left( -\right) ^{m+w}\hat{a}_{l-m}^{1-w}(k)],\quad\quad\quad  \tag{8} \\
\overset{\symbol{94}}{\Psi}\left( r\right) & =\sum\limits_{njmw}\left(
\left\langle r\upharpoonleft njmw\right\rangle \hat{d}_{njm}^{\frac{1}{2}%
w}+\left\langle r\upharpoonleft\overline{n}jmw\right\rangle \hat{d}_{njm}^{%
\frac{1}{2}w\dagger}\right) ,   \tag{9}
\end{align}
\newline
{}where the $\left\vert njmw\right\rangle $ and $\left\vert \overline{n}%
jmw\right\rangle $ form a complete set of quark and antiquark spinors with
angular momentum quantum numbers and spin-isospin quantum numbers $j,m,$ and
$w,$ respectively. The corresponding conjugate momentum fields have the
expansion,
\begin{align}
\hat{P}_{\sigma}\left( r\right) & =i\int\limits_{0}^{\infty}dkk^{2}\left[
\frac{W_{\sigma}\left( k\right) }{2\left( 2\pi\right) ^{3}}\right] ^{\frac{1%
}{2}}\left[ \hat{c}^{\dagger}\left( k\right) e^{-\mathbf{k}.\mathbf{r}}-\hat{%
c}\left( k\right) e^{^{+\mathbf{k}.\mathbf{r}}}\right] ,  \tag{10} \\
\hat{P}_{\pi}\left( r\right) & =i~\left[ \frac{2}{\pi}\right] ^{\frac {1}{2}%
}\int\limits_{0}^{\infty}dkk^{2}\left[ \frac{W_{\pi}\left( k\right) }{2}%
\right] ^{\frac{1}{2}}\sum\limits_{lmw}j_{l}\left( kr\right)
Y_{lm}^{\ast}\left( \Omega_{r}\right) [\hat{a}_{lm}^{1w\dagger}\left(
k\right) -  \notag \\
& \left( -\right) ^{m+w}\hat{a}_{l-m}^{1-w}\left( k\right) ].\quad
\quad\quad   \tag{11}
\end{align}
Here $\hat{c}\left( k\right) $ destroys a $\sigma$ with momentum $\mathbf{k}$
and frequency $W_{\sigma}\left( k\right) =\left( k^{2}+m_{\sigma}^{2}\right)
^{\frac{1}{2}}$ and $\hat{a}_{lm}^{1w}\left( k\right) $ destroys a pion with
momentum $\mathbf{k}$ and corresponding frequency $W_{\pi}\left( k\right)
=\left( k^{2}+m_{\pi}^{2}\right) ^{\frac{1}{2}}$ in the isospin-angular
momentum state $\{lm;tw\}$. For convenience, one constructs the
configuration space pion field functions needed for the subsequent
variational treatment by defining the alternative basis operators,
\begin{equation}
\hat{b}_{lm}^{1w}=\int dkk^{2}\zeta_{l}\left( k\right) \hat{a}%
_{lm}^{1w}\left( k\right) ,   \tag{12}
\end{equation}
where $\hat{a}_{lm}^{1w}\left( k\right) $ are basis operators which create a
free massive pion with isospin component $w$ and orbital angular momentum $%
(l,m)$, and $\zeta_{l}\left( k\right) $ is the variational function.
Considering this in configuration space, the pion field function [6] is
defined as
\begin{equation}
\Phi_{l}=\frac{1}{2\pi}\int\limits_{0}^{\infty}dkk^{2}\frac{\zeta_{l}\left(
k\right) }{W_{\pi}\left( k\right) ^{\frac{1}{2}}}j_{l}(r).   \tag{13}
\end{equation}

In the following, only the $l=1$ value is used and the angular momentum
label will be dropped. The Fock state for the nucleon is taken to be as in
Ref. $\left[ 6\right] $
% % % % % % % % % % % % % % % % % % % % % % % % % % % % % % % % % % % % % % % %

\section{\textbf{The Variational Principle}}

% % % % % % % % % % % % % % % % % % % % % % % % % % % % % % % % % % % % % % % %
In this section, the system of differential equations are presented,
following briefly the procedure of Refs. $\left[ 6,7\text{ and }8\right] $.
The objective of this section is to seek a minimum of the total baryon
energy, which is given by
\begin{equation}
E_{N}=\left\langle NT_{3}J_{z}\right\vert
\int\limits_{0}^{\infty}d^{3}r:H(r):\left\vert NT_{3}J_{z}\right\rangle ,
\tag{14}
\end{equation}
The field equations are obtained by minimizing the total energy of the
nucleon with respect to the variation of the fields, $\{u\left( r\right)
,v\left( r\right) ,\sigma\left( r\right) ,\Phi\left( r\right) \},$ as well
as the Fock-space parameters, $\left\{ \alpha,\beta,\gamma\right\} $,
subjected to the normalization conditions where $\alpha^{2}+\beta^{2}+%
\gamma^{2}=1$. The total energy of the system is written as
\begin{equation}
E_{N}=4\pi\int\limits_{0}^{\infty}drr^{2}\varepsilon_{N}\left( r\right) .
\tag{15}
\end{equation}
Writing the quark Dirac spinor as
\begin{equation}
\Psi_{\frac{1}{2}m}^{\frac{1}{2}w}\left( \mathbf{r}\right) =\left(
\begin{array}{c}
u\left( r\right) \\
v\left( r\right) \mathbf{\sigma}.\mathbf{\hat{r}}%
\end{array}
\right) \chi_{\frac{1}{2}m}\zeta^{\frac{1}{2}w},   \tag{16}
\end{equation}
the energy density is given by
\begin{align}
\varepsilon_{N}\left( r\right) & =\frac{1}{2}\left( \frac{d\sigma}{dr}%
\right) ^{2}+\frac{\lambda^{2}}{4}\left( \sigma^{2}\left( r\right)
-\nu^{2}\right) ^{2}-m_{\pi}^{2}f_{\pi}\sigma\left( r\right) +U_{0}  \notag
\\
- & \frac{1}{6\pi^{2}}(\frac{7\pi^{4}}{180}T^{4}+\frac{\pi^{2}}{6}%
T^{2}\mu^{2}+\frac{1}{12}\mu^{4})+6g^{2}(\frac{T^{2}}{12}+\frac{\mu^{2}}{%
4\pi^{2}})\sigma^{2}  \notag \\
& +3\left[ u\left( r\right) \left( \frac{dv}{dr}+\frac{2}{r}\upsilon\left(
r\right) \right) -\upsilon\left( r\right) \frac{du}{dr}+g\sigma\left(
r\right) \left( u^{2}\left( r\right) -\upsilon ^{2}\left( r\right) \right) %
\right]  \notag \\
& +\left( N_{\pi}+x\right) \left( \left( \frac{d\Phi}{dr}\right) ^{2}+\frac{2%
}{r^{2}}\Phi^{2}\left( r\right) \right) +\left( N_{\pi }-x\right)
\Phi_{p}^{2}\left( r\right) -  \notag \\
& \alpha\delta g\left( a+b\right) u\left( r\right) v\left( r\right)
\Phi\left( r\right) +\lambda^{2}[x^{2}+2xN_{\pi}+81\left(
\alpha^{2}a^{2}c^{2}+\left( \beta^{2}+\gamma^{2}\right) d^{2}\right)
]\Phi^{4}\left( r\right) +  \notag \\
& \lambda^{2}\left( N_{\pi}+x\right) \left( \sigma^{2}\left( r\right)
-v^{2}\right) \Phi^{2}\left( r\right) +12g^{2}(\frac{T^{2}}{12}+\frac {%
\mu^{2}}{4\pi^{2}})\left( N_{\pi}+x\right) \Phi(r)^{2}   \tag{17}
\end{align}
where $N_{\pi}$ is the average pion number%
\begin{equation}
N_{\pi}=9\left( \alpha^{2}a^{2}+\left( \beta^{2}+\gamma^{2}\right)
c^{2}\right) ,   \tag{18}
\end{equation}
and $\delta$ takes the following values for the nucleon quantum numbers:
\begin{equation}
\delta_{N}=\left( 5\beta+4\sqrt{2}\gamma\right) /\sqrt{3}.   \tag{19}
\end{equation}
The function $\Phi_{p}\left( r\right) $ is obtained from $\Phi\left(
r\right) $ by double folding:
\begin{align}
\Phi_{p}\left( r\right) & =\int\limits_{0}^{\infty}w\left( r,\acute {r}%
\right) \Phi\left( \mathbf{r}\right) r^{2}d\acute{r},  \tag{20} \\
w\left( r,\acute{r}\right) & =\frac{2}{\pi}\int\limits_{0}^{\infty}dkk^{2}w%
\left( k\right) j_{1}\left( kr\right) j_{1}\left( kr^{^{\prime}}\right) .
\tag{21}
\end{align}
For fixed $\alpha,\beta,$ and $\gamma,$ the stationary functional variations
are expressed by
\begin{equation}
\delta\left[ \int\limits_{0}^{\infty}drr^{2}(\varepsilon_{N}\left( r\right)
-3\epsilon\left( u^{2}\left( r\right) +v^{2}\left( r\right) \right)
-2k\Phi\Phi_{p}\left( r\right) )\right] =0,   \tag{22}
\end{equation}
where the parameter $k$ enforces the pion normalization condition,
\begin{equation}
8\pi\int\limits_{0}^{\infty}\Phi\left( r\right) \Phi_{p}\left( r\right)
r^{2}dr=1,   \tag{23}
\end{equation}
and $\epsilon$ fixes the quark normalization,
\begin{equation}
4\pi\int\limits_{0}^{\infty}(u^{2}\left( r\right) +v^{2}\left( r\right)
)r^{2}dr=1.   \tag{24}
\end{equation}

Minimizing the Hamiltonian yields the four nonlinear coupled differential
equations,
\begin{align}
\frac{du}{dr} & =-2(g\sigma+\epsilon)v(r)-\frac{1}{3}\alpha\delta\left(
a+b\right) g\Phi\left( r\right) u\left( r\right) ,  \tag{25} \\
\frac{dv}{dr} & =-\frac{2}{r}v(r)-2(g\sigma\left( r\right) -\epsilon
)u\left( r\right) +\frac{1}{3}\alpha\delta\left( a+b\right) g\Phi\left(
r\right) u\left( r\right) ,  \tag{26} \\
\frac{d^{2}\sigma}{dr^{2}} & =-\frac{2}{r}\frac{d\sigma}{dr}%
-m_{\pi}^{2}f_{\pi}+3g\left( u^{2}\left( r\right) -v^{2}\left( r\right)
\right) +2\lambda^{2}\left( N_{\pi}+x\right) \Phi^{2}\left( r\right)
\sigma\left( r\right) +  \notag \\
& +\lambda^{2}\left( \sigma^{2}\left( r\right) -\nu^{2}\right) \sigma\left(
r\right) +12g^{2}(\frac{T^{2}}{12}+\frac{\mu^{2}}{4\pi^{2}})\sigma\left(
r\right)  \tag{27} \\
\frac{d^{2}\Phi}{dr^{2}} & =-\frac{2}{r}\frac{d\Phi}{dr}+\frac{2}{r^{2}}%
\Phi\left( r\right) +\frac{1}{2}(1-\frac{x}{N_{\pi}})m_{\pi}^{2}\Phi +\frac{%
\lambda^{2}}{2}(1+\frac{x}{N_{\pi}})\left( \sigma^{2}\left( r\right)
-v^{2}\right) \Phi\left( r\right) -  \notag \\
& \frac{\alpha}{N_{\pi}}\left( a+b\right) g\delta u(r)v(r)+\frac {\lambda^{2}%
}{N_{\pi}}[x^{2}+2xN_{\pi}+81\left( \alpha^{2}a^{2}c^{2}+\left(
\beta^{2}+\gamma^{2}\right) d^{2}\right) ]\Phi^{3}\left( r\right)  \notag \\
& -\frac{k}{N_{\pi}}\Phi_{p}(r)+24g^{2}(\frac{T^{2}}{12}+\frac{\mu^{2}}{%
4\pi^{2}})\Phi(r)\left( 1+\frac{x}{N_{\pi}}\right) ,   \tag{28}
\end{align}
with the eigenvalues $\epsilon$ and $k.$ The above equations consist of two
quark equations for $u$ and $v\ $where $\sigma\left( r\right) $ and $%
\Phi\left( r\right) \ $appear as potentials and two Klein-Gordon equations
with $u\left( r\right) v\left( r\right) $ and $\left( u^{2}\left( r\right)
-v^{2}\left( r\right) \right) $ are source terms. The coefficients $a$, $b$,
and $c$ are functions in the coherence parameter $x$ and the field equations
are solved for the fixed coherence parameter $x$ and the fixed Fock-space
parameters ($\alpha,\beta,\gamma)$ as in Refs. [6,13].
% % % % % % % % % % % % % % % % % % % % % % % % % % % % % % % % % % % % % % % %

\section{The Nucleon Properties}

% % % % % % % % % % % % % % % % % % % % % % % % % % % % % % % % % % % % % % % %
The expectation value of the energy is minimized with respect to ($%
\alpha,\beta,\gamma)$ by diagonalizing the energy matrix%
\begin{equation}
\left[
\begin{array}{ccc}
H_{\alpha\alpha} & H_{\alpha\beta} & H_{\alpha\gamma} \\
H_{\alpha\beta} & H_{\beta\beta} & H_{\beta\gamma} \\
H_{\alpha\gamma} & H_{\beta\gamma} & H_{\gamma\gamma}%
\end{array}
\right] \left[
\begin{array}{c}
\alpha \\
\beta \\
\gamma%
\end{array}
\right] =E\left[
\begin{array}{c}
\alpha \\
\beta \\
\gamma%
\end{array}
\right] ,   \tag{29}
\end{equation}
where each $H$ entry of the matrix is related to a corresponding density as
follows:%
\begin{equation}
H_{\alpha\beta}=4\pi\dint r^{2}E_{\alpha\beta}(r)dr,   \tag{30}
\end{equation}
with similar definitions for the other entries. The functions for a nucleon
are%
\begin{align}
E_{\alpha\alpha} & =E_{0}(r)+18a^{2}\Phi_{P}^{2}++9a^{2}\lambda
^{2}(2x+9c^{2})\Phi^{4}(r)+9\lambda^{2}a^{2}\left( \sigma^{2}\left( r\right)
-\nu^{2}\right) \Phi^{2}(r)+  \notag \\
& 108a^{2}\Phi^{2}(r)g^{2}(\frac{T^{2}}{12}+\frac{\mu^{2}}{4\pi^{2}}),
\tag{31}
\end{align}
\begin{align}
E_{\beta\beta} &
=E_{0}(r)+18c^{2}\Phi_{P}^{2}+9\lambda^{2}(2xc^{2}+9d^{2})\Phi^{4}(r)+9%
\lambda^{2}c^{2}\left( \sigma^{2}\left( r\right) -\nu^{2}\right) \Phi^{2}(r)+
\notag \\
& 108c^{2}\Phi^{2}(r)g^{2}(\frac{T^{2}}{12}+\frac{\mu^{2}}{4\pi^{2}}),
\tag{32} \\
E_{\gamma\gamma} &
=E_{0}(r)+18c^{2}\Phi_{P}^{2}+9\lambda^{2}(2xc^{2}+9d^{2})\Phi^{4}(r)+9%
\lambda^{2}c^{2}\left( \sigma^{2}\left( r\right) -\nu^{2}\right) \Phi^{2}(r)+
\notag \\
& 108c^{2}\Phi^{2}(r)g^{2}(\frac{T^{2}}{12}+\frac{\mu^{2}}{4\pi^{2}}),
\tag{33} \\
E_{\alpha\beta} & =-2g(a+b)\Phi(r)u(r)v(r)\frac{2\sqrt{2}}{\sqrt{3}},
\tag{34} \\
E_{\alpha\gamma} & =-2g(a+b)\Phi(r)u(r)v(r)\frac{5}{\sqrt{3}},   \tag{35}
\end{align}
where:%
\begin{align}
E_{0}\left( r\right) & =\frac{1}{2}\left( \frac{d\sigma}{dr}\right)
^{2}+\lambda^{2}x^{2}\Phi^{4}\left( r\right) +3g\sigma\left( r\right) \left(
u^{2}\left( r\right) -\upsilon^{2}\left( r\right) \right)
-m_{\pi}^{2}f_{\pi}\sigma\left( r\right)  \notag \\
& +3[u\left( r\right) (\frac{dv}{dr}+\frac{2}{r}\upsilon\left( r\right)
-\upsilon\left( r\right) )\frac{du}{dr}]+\frac{\lambda^{2}}{4}\left(
\sigma^{2}\left( r\right) -\nu^{2}\right) ^{2}+\lambda^{2}x\left(
\sigma^{2}\left( r\right) -v^{2}\right) \Phi^{2}\left( r\right)  \notag \\
& +6g^{2}(\frac{T^{2}}{12}+\frac{\mu^{2}}{4\pi^{2}})\sigma^{2}+U_{0},
\tag{36}
\end{align}
where $U_{0}$ is the minimum of potential $U$ .
% % % % % % % % % % % % % % % % % % % % % % % % % % % % % % % % % % % % % % % %

\subsection{ Mass of the Nucleon}

% % % % % % % % % % % % % % % % % % % % % % % % % % % % % % % % % % % % % % % %
In this subsection, we calculate the total energy of the nucleon, which
consists of quark, sigma, pion, quark-sigma interaction, quark-pion
interaction, and meson static energy contributions. The nucleon mass was
derived as in Ref. [8]:%
\begin{equation}
(K.E)_{quark}=\int_{0}^{\infty}(g\sigma\rho_{s}(r)+\epsilon\rho_{w}(r)+g\pi%
\rho_{p}(r))r^{2}dr,   \tag{37}
\end{equation}
where $\rho_{s},\rho_{p},$ and $\rho_{w}$ are the quark scalar density,
pseudoscalar density, and vector density, respectively. Similarly, one can
find the meson kinetic contribution:%
\begin{align}
(K.E)_{sigma} & =\frac{1}{2}\int_{0}^{\infty}\sigma\left( r\right)
(-m_{\pi}^{2}f_{\pi}+3g\left( u^{2}\left( r\right) -v^{2}\left( r\right)
\right) +\lambda^{2}\left( N_{\pi}+x\right) \Phi^{2}\left( r\right)
\sigma\left( r\right) +  \notag \\
& +\lambda^{2}\left( \sigma^{2}\left( r\right) -\nu^{2}\right) \sigma\left(
r\right) +12g^{2}(\frac{T^{2}}{12}+\frac{\mu^{2}}{4\pi^{2}})\sigma)r^{2}dr
\tag{38}
\end{align}
\begin{align}
(K.E)_{pion} & =\frac{1}{2}\int_{0}^{\infty}\Phi\left( r\right) (\frac {2}{%
r^{2}}\Phi\left( r\right) +\frac{1}{2}(1-\frac{x}{N_{\pi}})m_{\pi}^{2}\Phi+%
\frac{\lambda^{2}}{2}(1+\frac{x}{N_{\pi}})\left( \sigma^{2}\left( r\right)
-v^{2}\right) \Phi\left( r\right) -  \notag \\
& \frac{\alpha}{4N_{\pi}}\left( a+b\right) g\delta u(r)v(r)+\frac {%
\lambda^{2}}{N_{\pi}}[x^{2}+2xN_{\pi}+81\left( \alpha^{2}a^{2}c^{2}+\left(
\beta^{2}+\gamma^{2}\right) d^{2}\right) ]\Phi^{3}\left( r\right)  \notag \\
& -\frac{k}{N_{\pi}}\Phi_{p}(r)+24g^{2}(\frac{T^{2}}{12}+\frac{\mu^{2}}{%
4\pi^{2}})\Phi(r)(1+\frac{x}{N_{\pi}}))r^{2}dr.   \tag{39}
\end{align}
The quark-meson interaction energy takes the form%
\begin{align}
E_{q-sigma} & =-\int_{0}^{\infty}g\sigma\rho_{s}(r)r^{2}dr  \tag{40} \\
E_{q-pion} & =-\int_{0}^{\infty}g\sigma\rho_{p}(r)r^{2}dr,   \tag{41}
\end{align}
and the meson-meson interaction energy is%
\begin{align}
E_{meson-meson} & =\int_{0}^{\infty}(\frac{\lambda^{2}}{4}\left( \hat{\sigma}%
^{2}+\mathbf{\hat{\pi}}^{2}-\nu^{2}\right) ^{2}-f_{\pi}m_{\pi }^{2}\hat{%
\sigma}  \notag \\
& \mathbf{-}\frac{6}{\pi^{2}}(\frac{7\pi^{4}}{180}T^{4}+\frac{\pi^{2}}{6}%
T^{2}\mu^{2}+\frac{1}{12}\mu^{4})+6g^{2}(\frac{T^{2}}{12}+\frac{\mu^{2}}{%
4\pi^{2}})\left( \sigma^{2}+\mathbf{\pi}^{2}\right) )r^{2}dr.   \tag{42}
\end{align}
% % % % % % % % % % % % % % % % % % % % % % % % % % % % % % % % % % % % % % % %

\section{Results and Discussion}

% % % % % % % % % % % % % % % % % % % % % % % % % % % % % % % % % % % % % % % %
In this section, we examine our calculations obtained from solving the
system of differential equations $\left( 25\rightarrow28\right) $. To solve
the system, we apply the modified COLSYS \ code at finite temperature and
chemical potential. The code is used before in many works such as in Refs. $%
\left[ 6\rightarrow8\right] .$ The initial parameters at zero temperature
and chemical potential are used as initial parameters at finite temperature
and chemical potential, where the mass of the sigma particle in the data
group ranges from 400 to 1200 MeV [26] and the coupling constant $%
g=3.5\rightarrow6$ are used as free parameters. Other parameters such as $%
m_{\pi}$ and $f_{\pi}$ are fixed from experimental data. The iteration
procedure is implemented as follows. For fixed values of $x$, $\alpha,\ \
\beta,$ and$\ \gamma$ , the differential equations with the corresponding
boundary conditions are solved until self-consistency is achieved.

The task of this section is to show the effect finite temperature and
chemical potential on nucleon properties. So, we divide the discussion into
two parts:

In the first part, the effect of finite chemical potential on the nucleon
properties is discussed as follows: In Fig. $\left( 1\right) ,$we displayed
the effect of finite chemical potential on the sigma, the pion, and the
quark fields at zero temperature and chemical potential in comparison with
their behavior at zero temperature and finite chemical potential. In Fig.
(1), we have plotted the above fields as functions of the distance $r$ . We
note that the sigma field increases with increasing distance $r$, in which
the sigma field takes a maximum value when $r\rightarrow\infty$. The
function $u(r)$ has maximum value at $r=0$ and then it decreases until it
reaches zero as $r\rightarrow\infty$. The function $v(r)$ has similar
behavior as the pion field. By increasing chemical potential up to 100 MeV,
we note that the qualitative features of the sigma and pion fields are
similar in the two cases where the fields shift to lower values in
comparison with the zero-temperature and chemical potential case. Hence, the
meson field is sensitive to the change of the chemical potential in
comparison with their behavior at zero temperature and chemical potential.
Therefore, the change in the mesonic contributions will be affected in the
observables of the nucleon as it will be\ explained next.

Now, we examine the change in the dynamic of the nucleon energy at finite
chemical potential. In Figs. (2, 3 and 4), we have plotted the kinetic
energy density of the quark, sigma and the pion fields and the interaction
between them at finite chemical potential and zero temperature in comparison
with their values at $\left( T=0,\mu=0\right) $. In Fig. 2, the kinetic
energies density of quark, sigma and pion fields are plotted as functions of
$r$ at finite chemical potential and zero-temperature in comparison with
their behavior at $\left( T=0,\mu=0\right) $. We note that the quark and
sigma kinetic energy density decreases with increasing $r$ up to $1$fm and
has a fixed value when $r$ tends to infinity. The pion kinetic energy
density increases up to $0.5$ fm and then it decreases with increasing
distance $r$ up to 1 fm and then it has a fixed value as $r\rightarrow\infty.
$ By increasing chemical potential up to 100 MeV, we note that the kinetic
energy density for each field is shifted to lower values, in particular, in
the range $r=0\rightarrow1$ fm. Thus, we deduce that the mesonic
contributions in these quantities are sensitive only at lower values of the
distance $r$.

In Fig. 3, the quark-meson density interaction is plotted as a function of $r
$. We \ note that the quark-sigma density interaction increases with
increasing $r$ up to 1 fm and then it has a fixed value when $r$ increases.
The quark-pion density has the smallest values in comparison with the
quark-sigma density, thus it appears as a fixed function. By increasing
chemical potential, we note that each field shifts to higher values at
smaller values of $r$. A similar situation takes place with respect to the
meson-meson density interaction that decreases with increasing distance $r$.
By increasing chemical potential, the curve is shifted to higher-values at
all values of $r$. The effect of the chemical potential clearly appears when
$r$ increases. Therefore, the meson-meson density interaction plays an
essential role at all values of $r$. Thus, this effect will act on the
mesonic contributions of the nucleon energy. \ The coherence parameter $x$
plays an important role in the present work. The nucleon mass is plotted as
a function of $x$ at $T=0$, $\mu=270$ MeV as in Fig. 5. We note that an
increase in the coherence parameter $x$ leads to an increase in the nucleon
mass due to the fact that the coherence parameter $x$ increases the pionic
contributions in the present model.

In Table \ref{tab1}, we calculate constituents of the nucleon energy at zero
temperature and chemical potential in comparison with the nucleon energy at
finite chemical potential. We note that the kinetic energy of the sigma,
pion and quark fields decrease with increasing chemical potential, while the
meson-meson interaction energy strongly increases. Therefore, we find that
the nucleon mass increases with increasing chemical potential. In addition,
by increasing chemical potential above 270 MeV, we find that the nucleon
mass decreases. %%%%%%%%%%%%%%%%%%%%%%%%%%%%%%%%%%%%%%%%%%%%%%%%%%

\begin{center}
\begin{table}[tbp]
\caption{The energy calculations of nucleon mass for value of coherence
parameter\textbf{\ }$x=1$\textbf{\ }at\textbf{\ }$m_{\protect\pi}=139.6$ MeV%
\textbf{, }$m_{\protect\sigma}=472$ MeV and $g=4.5.$ All observables in MeV.}%
\label{tab1} \vspace*{0.2cm}
\begin{tabular}{||l||l||l||}
\hline\hline
Quantity & $T=0$ and $\mu=0$ & $T=0$ and $\mu=270$ MeV \\ \hline\hline
Kinetic energy of quark & 463.390 & 312.538 \\ \hline\hline
Kinetic energy of sigma & 249.200 & 0.530 \\ \hline\hline
Kinetic energy of pion & 174.794 & 93.232 \\ \hline\hline
Quark-sigma interaction & 92.840 & -0.671 \\ \hline\hline
Quark-pion interaction & -1$\times10^{-4}$ & 0.745 \\ \hline\hline
Pion-sigma interaction & 136.47 & 1012 \\ \hline\hline
Nucleon Mass & 1069.835 & 1418.86 \\ \hline\hline
\end{tabular}%
\end{table}
\end{center}

%%%%%%%%%%%%%%%%%%%%%%%%%%%%%%%%%%%%%%%%%%%%%%%%%%

Here, we investigate other nucleon properties such as, the mean-square
radius of proton $\left\langle {\small r}^{2}\right\rangle _{p}$ and neutron
$\left\langle {\small r}^{2}\right\rangle _{n}$, the magnetic moment of
proton ${\small \mu}_{p}$ and neutron ${\small \mu}_{n}$, the coupling
constant $\frac{g_{A}}{g_{v}},$ and the pion-nucleon coupling constant $%
{\small g}_{\pi NN}^{G}$. In Table \ref{tab2}, we display the change in the
mesonic and quark contributions in these quantities when the chemical
potential is taken equal to $270$ MeV. We note that the quark contributions
in $\left\langle {\small r}^{2}\right\rangle _{p}$ strongly increases with
increasing chemical potential, while the mesonic contributions decreases
with increasing chemical potential. The total effect of quark and mesonic
contributions lead to an increase of $\left\langle {\small r}%
^{2}\right\rangle _{p}$ by increasing chemical potential. \ A similar
situation takes place with respect to $\left\langle {\small r}%
^{2}\right\rangle _{n}$. In the magnetic moment of proton, we note that the
mesonic and quark contributions decrease with increasing chemical potential.
Therefore, the magnetic moment of the proton decreases at finite chemical
potential in comparison with its value at zero temperature and chemical
potential. \ A similar situation takes place with respect to the magnetic
moment of neutron. The coupling constant $\frac{g_{A}}{g_{v}}$ and the
pion-nucleon constant ${\small g}_{\pi NN}^{G}$ play an essential role in
the present model. We note that two quantities decrease with increasing
chemical potential. %%%%%%%%%%%%%%%%%%%%%%%%%%%%%%%%%%%%%%%%%%%%%%%%%%

\begin{center}
\begin{table}[tbp]
\caption{The observables of the nucleon calculated for a value of the
coherence parameter\textbf{\ }$x=1$\textbf{\ }at\textbf{\ }$m_{\protect\pi%
}=139.6$ MeV\textbf{, }$m_{\protect\sigma}=472$ MeV and $g=4.5$.}\label{tab2}
\vspace*{0.2cm}
\begin{tabular}{||c||l||l||l||l||l||l||}
\hline\hline
& \multicolumn{3}{||l||}{$\ \ \ \ \ \ \ \ \ T=0$ and $\mu=0$} &
\multicolumn{3}{||l||}{$\ \ \ \ \ \ \ \ T=0$ and $\mu=270$ MeV} \\
\hline\hline
\multicolumn{1}{||l||}{\small Quantity} & {\small Quark} & {\small Meson} &
{\small Total} & {\small Quark} & {\small Meson} & {\small Total} \\
\hline\hline
\multicolumn{1}{||l||}{$\left\langle {\small r}^{2}\right\rangle _{p}$} &
{\small 7.217}$\times10^{-1}$ & {\small 2.8605}$\times10^{-2}$ & {\small %
0.750} & {\small 2.535} & {\small -2.584}$\times10^{-5}$ & {\small 2.535} \\
\hline\hline
\multicolumn{1}{||l||}{$\left\langle {\small r}^{2}\right\rangle _{n}$} &
{\small 2.232}$\times10^{-2}$ & {\small -2.8605}$\times10^{-2}$ & {\small %
-6.277}$\times10^{-3}$ & {\small -9.187}$\times10^{-6}$ & {\small 2.584}$%
\times10^{-5}$ & {\small 1.665}$\times10^{-5}$ \\ \hline\hline
\multicolumn{1}{||l||}{${\small \mu}_{p}$} & {\small 1.734} & {\small 0.175}
& {\small 1.909} & {\small 0.750} & {\small 6.659}$\times10^{-7}$ & {\small %
0.750} \\ \hline\hline
\multicolumn{1}{||l||}{${\small \mu}_{n}$} & {\small -1.262} & {\small -0.175%
} & {\small -1.437} & {\small 0.500} & {\small 6.659}$\times10^{-7}$ &
{\small -0.500} \\ \hline\hline
\multicolumn{1}{||l||}{$\frac{g_{A}}{g_{v}}$} & {\small 1.1624} & {\small %
0.371} & {\small 1.534} & {\small 0.296} & {\small 2.614}$\times10^{-7}$ &
{\small 0.296} \\ \hline\hline
\multicolumn{1}{||l||}{${\small g}_{\pi NN}^{G}$} & {\small 0.873} & {\small %
0.278} & {\small 1.151} & {\small 0.417} & {\small 1.90}$\times 10^{-7}$ &
{\small 0.417} \\ \hline\hline
\end{tabular}%
\end{table}
\end{center}

%%%%%%%%%%%%%%%%%%%%%%%%%%%%%%%%%%%%%%%%%%%%%%%%%%

Next, in the second part, we study the effect of finite temperature at zero
chemical potential in comparison with corresponding values at zero
temperature and chemical potential. We discuss the effect of finite
temperature on the dynamic of the fields. In Fig. 6, the sigma, the pion and
the components of the quark fields are plotted as functions of $r$ at finite
temperature in comparison with the behavior at zero temperature and chemical
potential. The sigma field increases with increasing distance $r$, in which
the sigma field has a maximum value as $r\rightarrow\infty$. Also, the pion
field takes the form of the P-wave function that increases up to 0.5 fm and
it decreases with increasing distance $r$. The components of the quark
fields $u\left( r\right) $ and $v\left( r\right) $ have maximum values at $%
r=0$ and tends to zero at infinity. The $v\left( r\right) $ component of the
quark filed takes a similar behavior to the pion field. By increasing
temperature, we note that all fields are shifted to lower values. Therefore,
the change in the thermal state leads to a change in the mesonic and quark
contributions which will have an effect on the dynamics of the fields.

In Fig. 7, the kinetic energy of the quark, sigma, and pion are plotted as
functions of $r$ \ at zero and finite temperature at vanishing chemical
potential. We note that the quark energy density decreases with increasing $r
$ up to 1 fm and then takes fixed values when $r$ $\rightarrow\infty.$ A
similar situation takes places with respect to the sigma kinetic energy
density, the difference between them appears at smallest values of $r$. The
kinetic energy density of the pion takes the form of P-wave function. By
increasing temperature, we find that the kinetic energy density of the
quark, the sigma and the pion fields are shifted to lower values. In Fig. 8,
the quark-meson density interaction is plotted. We note that the quark-sigma
density interaction increases with increasing distance $r$ up 1.2 fm and
then it has a fixed value when the distance $r$ \ increases. The quark-pion
density interaction has smallest values in comparison with its behavior at
finite temperature. By increasing temperature, we note the quark-meson
density interaction shifts to higher-values at smaller values of $r$. In
Fig. 9, the meson-meson density decreases with increasing $r.$ By increasing
temperature, we note that the curve shifts to higher values when $r$ \
increases. Therefore. the meson-meson density plays an essential role for
increasing nucleon energy due to an increase in the mesonic contributions at
larger values of the distance $r$.

In Table \ref{tab3}, we display the kinetic energy of the quark, the sigma
and the pion, the quark-meson interaction, and the meson-meson interaction
at $\ T=100$ MeV and $\mu=0$ in comparison with their values at zero
temperature and chemical potential.
%%%%%%%%%%%%%%%%%%%%%%%%%%%%%%%%%%%%%%%%%%%%%%%%%%

\begin{center}
\begin{table}[tbp]
\caption{The energy calculations of nucleon mass for value of coherence
parameter\textbf{\ }$x=1$\textbf{\ }at\textbf{\ }$m_{\protect\pi}=139.6$ MeV%
\textbf{, }$m_{\protect\sigma}=472$ MeV and $g=4.5.$ All observables in MeV.}%
\label{tab3} \vspace*{0.2cm}
\begin{tabular}{||l||l||l||}
\hline\hline
Quantity & $T=0$ and $\mu=0$ & $T=100$ MeV and $\mu=0$ \\ \hline\hline
Kinetic energy of quark & 463.390 & 210 \\ \hline\hline
Kinetic energy of sigma & 249.200 & -0.172 \\ \hline\hline
Kinetic energy of pion & 174.794 & 96.768 \\ \hline\hline
Quark-sigma interaction & 92.840 & 32.408 \\ \hline\hline
Quark-pion interaction & -1$\times10^{-4}$ & -0.003 \\ \hline\hline
Pion-sigma interaction & 136.47 & 1959.043 \\ \hline\hline
Nucleon Mass & 1069.835 & 2298.222 \\ \hline\hline
\end{tabular}%
\end{table}
\end{center}

%%%%%%%%%%%%%%%%%%%%%%%%%%%%%%%%%%%%%%%%%%%%%%%%%%

We note that the pion-sigma interaction increases strongly in comparison
with its value at $T=0$, $\mu=0$. Since, the meson-meson interaction is
sensitive to the changes at any value of the distance $r.$ The kinetic
energy of quark, sigma, and pion decrease with increasing temperature. The
strong change of the mesonic potential leads to an increase in the nucleon
mass up to $T=100$ MeV. This result is in agreement with the results of Ref.
$\left[ 21\right] $, in which the soliton energy increases with increasing
temperature at zero density using the Nambu-Jona- Lasinio model.
%%%%%%%%%%%%%%%%%%%%%%%%%%%%%%%%%%%%%%%%%%%%%%%%%%

\begin{center}
\begin{table}[tbp]
\caption{The observables of the nucleon calculated for coherence parameter%
\textbf{\ }$x=1$\textbf{\ }at\textbf{\ }$m_{\protect\pi}=139.6$ MeV\textbf{,
}$m_{\protect\sigma}=472$ MeV and $g=4.5$.}\label{tab4} \vspace*{0.2cm}
\begin{tabular}{||c||l||l||l||l||l||l||}
\hline\hline
& \multicolumn{3}{||l||}{$\ \ \ \ \ \ \ \ \ T=0$ and $\mu=0$} &
\multicolumn{3}{||l||}{$\ \ \ \ \ \ \ \ T=100$ MeV and $\mu=0$} \\
\hline\hline
\multicolumn{1}{||l||}{\small Quantity} & {\small Quark} & {\small Meson} &
{\small Total} & {\small Quark} & {\small Meson} & {\small Total} \\
\hline\hline
\multicolumn{1}{||l||}{$\left\langle {\small r}^{2}\right\rangle _{p}$} &
{\small 7.217}$\times10^{-1}$ & {\small 2.8605}$\times10^{-2}$ & {\small %
0.750} & {\small 2.974} & {\small -3.042}$\times10^{-5}$ & {\small 2.974} \\
\hline\hline
\multicolumn{1}{||l||}{$\left\langle {\small r}^{2}\right\rangle _{n}$} &
{\small 2.232}$\times10^{-2}$ & {\small -2.8605}$\times10^{-2}$ & {\small %
-6.277}$\times10^{-3}$ & {\small -1.089}$\times10^{-5}$ & {\small 3.0429}$%
\times10^{-5}$ & {\small 1.954}$\times10^{-5}$ \\ \hline\hline
\multicolumn{1}{||l||}{${\small \mu}_{p}$} & {\small 1.734} & {\small 0.175}
& {\small 1.909} & {\small 2.939} & {\small 7.462}$\times10^{-7}$ & {\small %
2.939} \\ \hline\hline
\multicolumn{1}{||l||}{${\small \mu}_{n}$} & {\small -1.262} & {\small -0.175%
} & {\small -1.437} & {\small -1.959} & {\small 7.462}$\times10^{-7}$ &
{\small -1.959} \\ \hline\hline
\multicolumn{1}{||l||}{$\frac{g_{A}}{g_{v}}$} & {\small 1.1624} & {\small %
0.371} & {\small 1.534} & {\small 5.046}$\times10^{-1}$ & {\small 3.76}$%
\times10^{-8}$ & {\small 0.5046} \\ \hline\hline
\multicolumn{1}{||l||}{${\small g}_{\pi NN}^{G}$} & {\small 0.873} & {\small %
0.278} & {\small 1.151} & {\small 3.711}$\times10^{-1}$ & {\small 7.57}$%
\times10^{-8}$ & {\small 0.3787} \\ \hline\hline
\end{tabular}%
\end{table}
\end{center}

%%%%%%%%%%%%%%%%%%%%%%%%%%%%%%%%%%%%%%%%%%%%%%%%%%

Next, we examine other nucleon properties given in Table \ref{tab4}: the
mean-square radius of proton $\left\langle {\small r}^{2}\right\rangle _{p}$%
and neutron $\left\langle {\small r}^{2}\right\rangle _{n}$, the magnetic
moment of proton ${\small \mu}_{p}$ and neutron ${\small \mu}_{n}$, the
coupling constant $\frac{g_{A}}{g_{v}}$, and the pion-nucleon coupling
constant ${\small g}_{\pi NN}^{G}.$ We note that the $\left\langle {\small r}%
^{2}\right\rangle _{p}$ increases with increasing temperature, where the
mesonic contributions decrease in comparison with its value at $T=0$ and $%
\mu=0,$ while the quark contributions increase three times in comparison
with its value at $T=0$ and $\mu=0.$ Thus the quark contributions have more
effect on the value of the mean-square radius of proton. Therefore, we note
that $\left\langle {\small r}^{2}\right\rangle _{p}$ increases with
increasing temperature. A similar situation takes place with respect to $%
\left\langle {\small r}^{2}\right\rangle _{n}$. \ An increase in $%
\left\langle {\small r}^{2}\right\rangle _{p}$ is noted in Refs. $\left[
12,13,14\right] $ \ with increasing temperature, in which the quark sigma
model is applied in the mean-field approximation. The magnetic moment of the
proton and neutron increase with increasing temperature. We note that the
quark contributions strongly increase with increasing temperature, while the
mesonic contributions decrease with increasing temperature. The coupling
constant $\frac{g_{A}}{g_{v}}$ and the pion-nucleon coupling constant $%
{\small g}_{\pi NN}^{G}$ are also calculated. We note that the quark and
mesonic contributions decrease with increasing temperature. Therefore, we
find that $\frac{g_{A}}{g_{v}}$ and ${\small g}_{\pi NN}^{G}$ decrease with
increasing temperature. In the present work, we find that the nucleon mass
and mean-square radius of the proton increase with increasing temperature,
while the pion-nucleon coupling constant decreases with increasing
temperature. This indicate that a deconfinment phase is satisfied at higher
values of the temperature and in agreement with QCD finite energy sum rule $%
\left[ 13\right] $.
% % % % % % % % % % % % % % % % % % % % % % % % % % % % % % % % % % % % % % % %

\section{Summary and Conclusions}

% % % % % % % % % % % % % % % % % % % % % % % % % % % % % % % % % % % % % % % %
In the previous section, we gave qualitative and quantitative description of
the meson fields and the nucleon properties at finite temperature and
chemical potential. So, we need to show clearly the novelty in this work. We
gave in the section of introduction that most of nucleon properties at
finite temperature and chemical potential are carried out in the framework
of chiral sigma model and NJL model in the mean-field approximation. The
coherent-pair approximation (CPA) is applied in the chiral quark model at
finite temperature only in the previous work $\left[ 14\right] $. \ CPA is a
powerful nonperturbative method to go beyond the mean-field approximation by
fully taking thermal and quantum fluctuations into account $\left[ 14\right]
$. In addition, one avoids assumptions like the hedgehog structure of the
quark and pion fields that are taken in many works at finite temperature and
density such as in Ref. $\left[ 17\right] $ and the references therein. So
far no attempt has been made to include the finite chemical potential by
using the coherent-pair approximation.

The qualitative features\ of sigma and pion fields are in agreement with
mean-field approximation results at finite temperature and chemical
potential. In addition, the sigma and pion fields are sensitive to change of
the temperature and chemical potential. The pion-nucleon coupling constant
and pion decay constant decreases with increasing temperature and chemical
potential, which indicate that a deconfinement phase transition is satisfied
in the present work at higher values of the temperature. The mean-square
radius of proton $\left\langle {\small r}^{2}\right\rangle _{p}$ and neutron
$\left\langle {\small r}^{2}\right\rangle _{n}$, the magnetic moment of
proton ${\small \mu}_{p}$ and neutron ${\small \mu}_{n}$ are investigated at
finite chemical potential. The qualitative agreement of present results are
discussed in comparison with other works. The effect of coherence of the
parameter $x$ on the nucleon mass is studied.

In the coherent-pair approximation, we conclude that the nucleon properties
are strong changed by including finite chemical potential. The coherence
parameter $x$ plays an important for changing the nucleon mass at finite
chemical potential, since an increase in x leads to increasing in the pionic
contributions in the energy of nucleon at finite chemical potential.

We hope to extend this work to include external magnetic field which will
give better understanding into the complex vacuum structure of quantum
chromodynamic (QCD) theory which describes relevant features of particle
physics in the early universe and the neutron stars.
%%%%%%%%%%%%%%%%%%%%%%%%%%%%%%%

\newpage

%%%%%%%%%%%%%%%%%%%%%%%%%%%%%%%%%%%%%%%%%%%%%%%%
\begin{figure}
\centerline{\psfig{figure=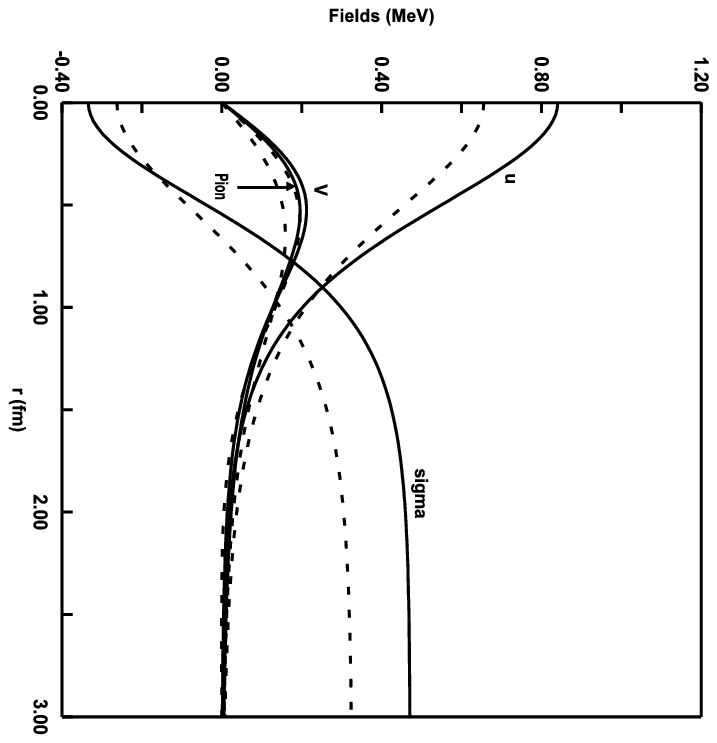,height=3in,width=4in,angle=-90}}
\caption{Sigma, pion, and the components $(u(r),v(r))$ of the quark fields
are plotted as functions of the radial distance r, where the continuous
curves are for $T=0$ and $\protect\mu=0$ and the dashed curves are for $T=0$
and $\protect\mu=100$ MeV.}
\label{fig1fig}
\end{figure}